\documentclass{article}
\usepackage{amsmath}
\usepackage{amsfonts}
\numberwithin{equation}{section}
\newcommand{\R}{\mathbb R}
\newcommand{\E}{\mathbb E}
\newcommand{\G}{\mathcal{G}}

\begin{document}
\begin{titlepage}
\begin{center}
{\Large \bf Super-Matrix KdV and\\ Super-Generalized NS Equations \\
from\\Self-Dual Yang-Mills Systems with\\Supergauge Groups}
\vskip 5em
{\large  J. LaChapelle and M. Legar\'e}\\
\vskip 1em
Department of Mathematical Sciences\\
University of Alberta\\
Edmonton, Alberta, Canada, T6G 2G1
\end{center}
\vskip 5em

\begin{abstract}
\sloppy{Super-matrix KdV and super-generalized non-linear
Schr\"{o}dinger
equations are shown to arise from a symmetry reduction of ordinary
self-dual Yang-Mills equations with supergauge groups. PACS:
11.15.-q, 11.30.Pb}
\end{abstract}
\end{titlepage}

\pagebreak

\section{Introduction}
It has been known for some time that many completely integrable
systems in $1+1$ dimensions can be obtained by symmetry reduction
of self-dual Yang-Mills (SDYM) field equations in the
four-dimensional Euclidean space $\E^{4}$ or the $(2,2)$
pseudo-Euclidean
space $\E^{(2,2)}$, which corresponds to $\R^{4}$ endowed with the
diagonal
metric : $diag(+1,+1,-1,-1)$ \cite{AC,W1}. Super-extensions of known
integrable systems in $1+1$ dimensions have also
been obtained from supersymmetric SDYM field equations by means of
symmetry
reduction \cite{GN,GMRT}.

Some insight into this relationship can be achieved by examining
the linear system associated with the non-linear SDYM equations.
Both the SDYM and its associated linear system are invariant under
conformal as well as gauge transformations. Therefore, the
integrability conditions of the linear system reduced by a subgroup
of the conformal group coincide with the similarly reduced SDYM
field equations. This fact makes the linear system invaluable in
the study of the reduction of the SDYM field equations to known
integrable systems by means of symmetry. Indeed, to show that the
SDYM field equations give rise to a known integrable system, one
only needs to find the appropriate reduction symmetry, gauge group,
and linear system. Accordingly, a linear system for an integrable
system can be revealed if the reduction steps from the SDYM
equations to the integrable system are applied to the SDYM linear
system.

Linear systems known as Lax pairs have also been useful in the
study of \mbox{(super-)} integrable systems in $1+1$ dimensions and
their associated hierarchies \cite{AC}. Lax pairs constructed from
general techniques (e.g. refs. \cite{DS,ADR,MP}) can sometimes be
utilized to help find the symmetry reductions of SDYM equations
which lead to known (super-) integrable systems. In fact, ref.
\cite{L1} showed that a super-generalized non-linear
Schr\"{o}dinger (super-GNS) equations can be obtained from symmetry
reduction of SDYM systems with gauge group $SL(2/1)$ using the
methods of ref. \cite{ADR}.

However, in general the Lax pairs may involve a (spectral)
parameter, $\lambda$, of order $O(\lambda^{3})$ or higher. This
will preclude application of these Lax pairs to SDYM systems
reduced by translation because the latter permit a spectral
parameter of order at most $O(\lambda^{2})$. For instance, ref.
\cite{MP} gives a Lax pair for the super-KdV equation, but it
contains a spectral parameter of order $O(\lambda^{3})$.

There is an alternate Lax pair formulation for the ordinary (matrix)
KdV equation with spectral parameter of order $O(\lambda^{2})$
\cite{BP,IP1,IP2}. This suggests that it might be possible to find
an order $O(\lambda^{2})$ linear system for super-(matrix) KdV as
well. It is indeed possible, and in subsection \ref{Super-Matrix
KdV} we will exhibit linear systems with a spectral parameter
$O(\lambda^{2})$ which give rise to super-KdV and super-matrix KdV
equations.

Consequently, the super-(matrix) KdV as well as the super-GNS
equations can be obtained from a symmetry reduction of SDYM systems
with a supergauge group.  This could be seen as a supersymmetric
extension of the SDYM reductions to the KdV and NS equations
presented in \cite{MS}. It is noteworthy that \emph{after} symmetry
reduction the ordinary SDYM equations acquire a supersymmetry
associated with spatial translations.

A theory-based procedure for finding linear systems associated with
the super-matrix KdV is not shown here. Our results were obtained
by considering the most general possible supergauge fields along
with some guidance from the non-supersymmetric case.

\section{SDYM Linear System}\label{Associated Linear System}

We begin with a principal bundle $P(B,\G)$. The base space $B$ is
$\E^{(2,2)}$, and the structure group $\G$ is a supergroup of even
dimension $m$ and odd dimension $n$. The gauge potential
$\boldsymbol A_{\mu}$, where $\mu\in\{1,\dots,4\}$, is Lie
superalgebra (cf : \cite{Ka,Co}) valued:
\begin{equation}\label{gauge potential}
\boldsymbol A_{\mu}(x)=A_{\mu}^{a}(x)\mathbf M_{a}+\xi_{\mu}^{\alpha}(x)
\mathbf N_{\alpha}.
\end{equation}
 Here $x\in B$, and $\{\mathbf M_{a},\ a\in\{1,\dots,m\}\}$
represent the even and $\{\mathbf N_{\alpha},\
\alpha\in\{1,\dots,n\}\}$ the odd basis elements of the Lie
superalgebra. We will restrict our attention to gauge potentials
which are pure, i.e.,\ either anti-commuting ($a$-type) or commuting
($c$-type).

The SDYM field equations, $\boldsymbol F=*\boldsymbol F$, are a set
of nonlinear partial differential equations in terms of the gauge
fields. Here $\boldsymbol F =d\omega + \omega\wedge\omega$ is the
curvature 2-form associated with the 1-form $\omega = \boldsymbol
A_\mu
\theta^\mu$ where $\{\theta^\mu\}$ is a basis of $T^\ast B$ and $*$ is
the
Hodge operator or duality transformation associated with the $(2,2)$
pseudo-Euclidean metric. It is recalled that the self-dual field
equations considered involve gauge fields with both bosonic and
fermionic variables coupled by the field equations.

There exists a linear system of partial differential equations
whose integrability condition reproduces the non-linear SDYM
equations. The linear system can be written
\begin{equation}\label{linear system}
  \begin{cases}
  (D_{1}+iD_{2}+\lambda (D_{3}-iD_{4}))\Psi(x,\lambda,\bar\lambda)=0\\
  (D_{3}+iD_{4}+\lambda (D_{1}-iD_{2}))\Psi(x,\lambda,\bar\lambda)=0
 \end{cases}
\end{equation}
where $D_{\mu}:=\partial_{\mu}+\boldsymbol A_{\mu}(x)$,
$\lambda\in\mathbb H^2$-sheet, and $\partial_{\bar\lambda}\Psi=0$.
The nature of $\Psi$, which can be inferred from the twistor
construction of the linear system \cite{WW}, has no direct bearing
on the developments in this letter. In other words, for our purposes
the specific nature of $\Psi$, which could be a multiplet of scalar
fields with values in the adjoint representation of the Lie
(super)algebra, does not matter and \eqref{linear system} can be
viewed as operator equations.

Following \cite{IP1,IP2}, we introduce null coordinates on
$\E^{2,2}$ defined by
\begin{align} \label{coor trans}
  t & =2^{-\frac{1}{2}}(x_{2}-x_{4})& y &
=2^{-\frac{1}{2}}(x_{1}-x_{3})  \\
  u & =2^{-\frac{1}{2}}(x_{2}+x_{4})& z &
=2^{-\frac{1}{2}}(x_{1}+x_{3}). \nonumber
\end{align}
In these coordinates, the covariant derivatives become
  \begin{align} \label{covariant derivatives}
    D_{t} & = 2^{-\frac{1}{2}}(D_{2}-D_{4}) &  D_{y} &
=2^{-\frac{1}{2}}(D_{1}-D_{3})\\
    D_{u} & =2^{-\frac{1}{2}}(D_{2}+D_{4})& D_{z} &
=2^{-\frac{1}{2}}(D_{1}+D_{3}). \nonumber
  \end{align}

Consequently, the linear system \eqref{linear system} on
$B=\E^{(2,2)}$ decouples:
\begin{equation}\label{new linear system}
  \begin{cases}
  (D_{z}+\omega D_{u})\Psi(x,\omega,\bar\omega)&\!\!\!=0\\
  (D_{t}-\omega D_{y})\Psi(x,\omega,\bar\omega)&\!\!\!=0
 \end{cases}
\end{equation}
where $\omega:=\tfrac{i(1-\lambda)}{(1+\lambda)}$. The holomorphy
condition on $\Psi$ becomes simply: $\partial_{\bar\omega}\Psi=0$.

\section{Reduced Linear System}\label{Reduced Linear System}
The SDYM field equations and their associated linear system are
invariant under both conformal transformations on the base space
$B$ as well as supergauge transformations. The conformal invariance
can be used to reduce the SDYM and its associated linear system
with respect to any subgroup, $G$, of the conformal group. Loosely
speaking, it entails rewriting
\eqref{new linear system} in terms of orbit and invariant
coordinates of the $G$-action and inserting $G$-invariant gauge
fields and a $G$-invariant $\Psi$.\footnote{One may consult refs.
\cite{FM,HSV,JM,HNO,Ol,Wi,L2,LP} for a complete description of the
general symmetry reduction method including details concerning
general invariance conditions on gauge fields and the holomorphic
nature of $\Psi$.}

We are particularly interested in the subgroup
$\tilde{G}=\{P_{u},P_{y}-P_{z}\}$, where $P_X$ denotes the generator of
translations along the $X$ coordinates, e.g. $P_X =
\dfrac{\partial}{\partial X}$ as a (vector field) representation.
Reducing
\eqref{new linear system} by $\tilde{G}$ leads to the $O(\omega^{2})$
linear system
\cite{IP1,IP2}
\begin{equation}\label{reduced linear system}
  \begin{cases}
  (\partial_{t}+\boldsymbol A_{t}(t,x)+\omega [\boldsymbol A_{z}(t,x)-
\boldsymbol A_{y}(t,x)]+\omega^{2}\boldsymbol A_{u}(t,x))\Psi=0\\
(\partial_{x}+\boldsymbol A_{z}(t,x)+\omega \boldsymbol
A_{u}(t,x))\Psi=0
 \end{cases}
\end{equation}
where $x:=(y+z)$. The integrability condition of \eqref{reduced
linear system} yields
\begin{eqnarray}\label{integrability}
  \partial_{x}\boldsymbol A_{u}+[\boldsymbol A_{y},\boldsymbol
  A_{u}]=0\;\;\;\; & & O(\omega^{2})\\
  \partial_{t}\boldsymbol A_{u}-\partial_{x}(\boldsymbol
A_{z}-\boldsymbol
A_{y})+[\boldsymbol A_{z},\boldsymbol A_{y}]+[\boldsymbol
A_{t},\boldsymbol A_{u}]=0\;\;\;\; & & O(\omega^{1})\nonumber\\
  \partial_{t}\boldsymbol A_{z}-\partial_{x}\boldsymbol
A_{t}+[\boldsymbol
A_{t},\boldsymbol A_{z}]=0\;\;\;\; & & O(\omega^{0}).\nonumber
\end{eqnarray}
Equations \eqref{integrability} coincide with the similarly reduced
SDYM field equations.

\subsection{Example: Super-Matrix KdV Equations}\label{Super-Matrix KdV}
The compatibility condition \eqref{integrability} for the reduced
linear system can be used to exhibit a super-extension of the
matrix KdV equation. We consider a SDYM theory with supergauge
group $\G=GL(2n/n)$ and choose the following gauge fields:
 \begin{align}\label{matrix KdV gauge fields}
 \boldsymbol A_{u}(t,x)=\begin{pmatrix}
        \mathbf 0_{n} & \mathbf 0_{n} & \mathbf 0_{n} \notag \\
         \mathbf {-1}_{n} & \mathbf 0_{n} & \mathbf 0_{n} \notag \\
         \mathbf 0_{n} & \mathbf 0_{n} & \mathbf 0_{n} \notag
         \end{pmatrix}, & &
 \boldsymbol A_{z}(t,x)=\begin{pmatrix}
         \mathbf 0_{n} & \mathbf 0_{n} & \mathbf 0_{n} \notag \\
         \mathbf u_{n}(t,x) & \mathbf 0_{n} & \mathbf 0_{n} \notag \\
         \mathbf 0_{n} & \mathbf 0_{n} & \theta
\boldsymbol{\varphi}_{n}(t,x) \notag
        \end{pmatrix}, \\
\\
\boldsymbol A_{z-y}(t,x)=\begin{pmatrix}
         \mathbf 0_{n} & \mathbf 0_{n} & \mathbf 0_{n} \notag \\
         \mathbf 0_{n} & \mathbf 0_{n} & \theta\mathbf 1_{n} \notag \\
         \theta\mathbf 1_{n} & \mathbf 0_{n} & \mathbf 0_{n} \notag
         \end{pmatrix}, & &
\boldsymbol A_{t}(t,x)=\begin{pmatrix}
         \mathbf 0_{n} & \mathbf 0_{n} & \mathbf 0_{n} \notag\\
         \mathbf a_{21}(t,x) & \mathbf 0_{n} & \mathbf 0_{n} \notag\\
         \mathbf 0_{n} & \mathbf 0_{n} & \theta\mathbf a_{33}(t,x)
\notag
         \end{pmatrix}.
 \end{align}
Here $\mathbf a_{21}\!:=3\mathbf u^{2}_{n}+\mathbf
u_{n,xx}-3\boldsymbol{\varphi}_{n}\boldsymbol{\varphi}_{n,x}$,
\  $\mathbf
a_{33}\!:=\boldsymbol{\varphi}_{n,xx}+3\boldsymbol{\varphi}_{n}\mathbf
u_{n}$,
\ $\boldsymbol A_{z-y}\!:=\boldsymbol A_{z}-\boldsymbol A_{y}$, $\mathbf
u_{n}$ is even, $\boldsymbol{\varphi}_{n}$ is odd, and $\theta$ is
an odd parameter. The $n$ subscript indicates an $n\times n$ matrix.

Substituting \eqref{matrix KdV gauge fields} into
\eqref{integrability} yields super-matrix KdV equations:
\begin{align}\label{matrix KdV}
   \mathbf u_{n,t}&=
        \mathbf
        u_{n,xxx}+3(\mathbf u_{n} \mathbf u_{n,x}+\mathbf
u_{n,x}\mathbf u_{n})-3(\boldsymbol {\varphi}_{n,x}\boldsymbol
{\varphi}_{n,x}+\boldsymbol {\varphi}_{n}\boldsymbol
        {\varphi}_{n,xx})\\
   \boldsymbol{\varphi}_{n,t}&=
       \boldsymbol{\varphi}_{n,xxx}+3(\boldsymbol
{\varphi}_{n}\mathbf u_{n,x}+
        \boldsymbol{\varphi}_{n,x}\mathbf u_{n}).\nonumber
\end{align} These super-matrix KdV equations reduce to the ordinary
matrix KdV equation \cite{WK} when $\boldsymbol\varphi_{n}=0$, and
they are invariant under the supersymmetry transformations:
\begin{align}\label{susy trans}
  \delta_{\varepsilon}\mathbf
  u_{n}=\varepsilon\boldsymbol\varphi_{n,x},\;\;\;\; &
  \delta_{\varepsilon}\boldsymbol\varphi_{n}=\varepsilon\mathbf
  u_{n}.
\end{align} Moreover, when $n=1$, \eqref{matrix KdV} reduces to the
super-KdV equations \cite{M}:
\begin{align}\label{super KdV}
   u_{,t}&= u_{,xxx}+6u u_{,x}-3\varphi\varphi_{,xx}\\
   \varphi_{,t}&=
   \varphi_{,xxx}+3(u_{,x}\varphi+u\varphi_{,x}).\nonumber
\end{align}

Linear systems for the super-matrix KdV equations can also be
constructed for the gauge groups $SL(2n/n)$ and $SL(n/n)$. Noticing
that the right-hand sides of
\eqref{matrix KdV} are total partial derivatives, the construction
is rather simple. For $SL(2n/n)$ we have
\begin{align}\label{medium matrix KdV}
 \boldsymbol A_{u}(t,x)=\begin{pmatrix}
        \mathbf 0_{n} & \mathbf 0_{n} & \mathbf 0_{n} \notag \\
         \mathbf {-1}_{n} & \mathbf 0_{n} & \mathbf 0_{n} \notag \\
         \mathbf 0_{n} & \mathbf 0_{n} & \mathbf 0_{n} \notag
         \end{pmatrix}, & &
 \boldsymbol A_{z}(t,x)=\begin{pmatrix}
         \mathbf 0_{n} & \mathbf 0_{n} & \mathbf 0_{n} \notag \\
         \mathbf u_{n}(t,x)+\theta \boldsymbol{\varphi}_{n}(t,x) &
\mathbf
0_{n} & \mathbf 0_{n} \notag \\
         \mathbf 0_{n} & \mathbf 0_{n} &  \mathbf 0_{n}\notag
        \end{pmatrix}, \\
\\
\boldsymbol A_{z-y}(t,x)=\begin{pmatrix}
         \mathbf 0_{n} & \mathbf 0_{n} & \mathbf 0_{n} \notag \\
         \mathbf 0_{n} & \mathbf 0_{n} & \mathbf 0_{n} \notag \\
         \mathbf 0_{n} & \mathbf 0_{n} & \mathbf 0_{n} \notag
         \end{pmatrix}, & &
\boldsymbol A_{t}(t,x)=\begin{pmatrix}
         \mathbf 0_{n} & \mathbf 0_{n} & \mathbf 0_{n} \notag\\
         \mathbf a_{21}(t,x)+\theta\mathbf a_{33}(t,x) & \mathbf 0_{n} &
\mathbf 0_{n} \notag\\
         \mathbf 0_{n} & \mathbf 0_{n} & \mathbf 0_{n} \notag
         \end{pmatrix}.
 \end{align}

The case $SL(n/n)$ is obtained from
\eqref{medium matrix KdV} by deleting the third rows and columns, and
adjusting $\boldsymbol A_{z}$ and $\boldsymbol A_{t}$ by moving the
odd parameter $\theta$ over to the even components. When $n=1$, one
can delete the first rows and second columns to obtain a linear
system for the super-KdV equations.

\subsection{Example: Super-Generalized NS Equations}

The supersymmetric version of the generalized non-linear
Schr\"odinger equations related to the symmetric space $SO(3)/SO(2)$
presented in refs. \cite{ADR,AR} can also be obtained by the same
reduction
with gauge group $SL(2/1)$. Let us recall the invariant gauge field
components of the nonlinear system
\eqref{reduced linear system} \cite{L1}:

\begin{eqnarray}\label{GNS gauge fields}
 \boldsymbol A_{u}(t,x)&=&\begin{pmatrix}
       -1 & 0 & 0\notag \\
         0 & 0 & 0 \notag \\
         0 & 0 & -1\notag
         \end{pmatrix},\\
 \boldsymbol A_{z}(t,x)&=& \boldsymbol A_{z-y}(t,x)=\begin{pmatrix}
         0 &\Phi & 0 \notag \\
         -D\xi+(\partial_x^{-1}\Phi\xi)\xi&  0 & \xi \notag\\
         0 &D\Phi+(\partial_x^{-1}\Phi\xi)\Phi  & 0 \notag
\end{pmatrix},\\ \notag\\
\boldsymbol A_{t}(t,x)&=& - \partial_x^{-1}([\boldsymbol
A_z,\mathcal{R}(\boldsymbol A_z)])
 + \mathcal{R}(\boldsymbol A_z),
\end{eqnarray}
where $\mathcal{R} = ad_{\boldsymbol
A_u}(\partial_x-ad_{\boldsymbol
A_z}\,\partial_x^{-1}\,ad_{\boldsymbol A_z})$, the bosonic field
$\Phi=\phi_1+\psi_1\theta$, the fermionic variable
$\xi=\psi_2+\phi_2\theta$, and $D=\partial_\theta +
\theta\partial_x$.

The corresponding reduced SDYM equations can be written as \cite{ADR,AR}
:
\begin{eqnarray}
\phi_{1,t} - \phi_{1,xx}+2\phi_1\psi_2\psi_1-2\phi_1^2\phi_2 &=&
0,\nonumber\\
\psi_{1,t} - \psi_{1,xx} - 2\phi_1\phi_{1,x}\psi_2 - 2\phi_1^2\psi_{2,x}
-
2\phi_1\phi_2\psi_1 &=& 0,\nonumber\\
\psi_{2,t} + \psi_{2,xx} + 2\psi_2\phi_1\phi_2 &=& 0,\nonumber\\
\phi_{2,t} + \phi_{2,xx} - 2\psi_2\psi_1\phi_2 + 2\phi_1\phi_2^2 -
2\phi_1\psi_2\psi_{2,x} &=& 0,
\end{eqnarray}
They are left invariant under the supersymmetry transformations
\cite{ADR,AR} :
\begin{equation}
\delta_{\varepsilon}\phi_1 = \varepsilon \psi_1 ,\;
\delta_{\varepsilon}\phi_2 = \varepsilon \psi_{2,x},\;
\delta_{\varepsilon}\psi_1 = \varepsilon \phi_{1,x},\;
\delta_{\varepsilon}\psi_2 = \varepsilon \phi_2.
\end{equation}

\section{Summary}

In this letter, the SDYM equations with $GL(m/n)$ supergauge group
reduced under null and space translations have allowed to recover,
with the addition of specific constraints on the gauge field
components, a supersymmetric version of the matrix KdV equations.
An outline of a similar SDYM reduction to a supersymmetric
formulation of the GNS equations has been given for the gauge group
$SL(2/1)$. A possible future research direction would be to
investigate the reduction, by different subgroups of the conformal
group acting on either $\E^4$ or $\E^{(2,2)}$, of the SDYM
equations and their linear system in order to derive known
\mbox{(super-)} integrable systems or new supersymmetric versions
of integrable equations. Since the reduced systems obtained via the
above procedure are not necessarily supersymmetric, a method
allowing the determination of (all) the supersymmetries of the
reduced equations would be useful. In view of recent results on
cohomological quantum field theories (see for example : refs
\cite{BKS,BT,AFSOL}), a similar approach (symmetry reduction) could be
applied to higher-dimensional (or generalized) self-dual Yang-Mills
equations \cite{CDFN,W2,IP3,I1,IP4,AOL,BL}.

\section{Note Added in Proof}
After submission of this work, ref. \cite{KS}, which uses a
different formulation to obtain $N=2$ supersymmetric matrix
$(1,s)$-KdV hierarchies, was found on the hep-th archives.

\section{Acknowledgements}

J.L. acknowledges a postdoctoral fellowship supported through funds
from a grant from the National Sciences and Engineering Research
Council (NSERC) of Canada, and M.L. acknowledges for the support of
this work a grant from (NSERC) of Canada.

\end{document}